# Counting Individual Electrons on Liquid Helium

G. Papageorgiou[1], P. Glasson[1], K. Harrabi[1], V.Antonov[1], E.Collin[2], P.Fozooni[1],

P.G.Frayne[1], M.J.Lea[1], Y.Mukharsky[2] and D.G.Rees[1].

[1]Department of Physics, Royal Holloway, University of London, Egham, Surrey,

TW20 0EX, UK.

[2]CEA, Saclay, France.

29 April 2004

We show that small numbers of electrons, including a single electron, can be held in a novel electrostatic trap above the surface of superfluid helium. A potential well is created using microfabricated electrodes in a 5 μm diameter pool of helium. Electrons are injected into the trap from an electron reservoir on a helium microchannel. They are individually detected using a superconducting single-electron transistor (SET) as an electrometer. A Coulomb staircase is observed as electrons leave the trap one-by-one until the trap is empty. A design for a scalable quantum information processor using arrays of electron traps is presented.

73.20.-r, 73.50.Gr

The smallest element for electronic quantum information storage would be a single electron in a quantum dot. Two quantum states form the basis |0> and |1> for a quantum bit, or qubit, while interacting qubits enable quantum information processing (QIP)[1]. Trapping, control and detection of single electrons are major technical challenges. Single electrons, and a positron named Priscilla[2], have been held in a Penning trap and detected by resonant interactions. In the solid state, electrons have been trapped in a chain of tunnel junctions[3]; one-electron quantum dots have been made in self-assembled structures[4], etched vertical pillars[5] and lateral quantum dots[6]; single[7] and double[8] dots have incorporated a quantum point contact electrometer. Here we show that surface-state electrons can be held in a novel electrostatic trap on a pool of superfluid helium, 5 μm in diameter and 0.8 μm deep. Individual localised electrons are detected and counted by a single-electron transistor (SET) electrometer beneath the helium. This leads to a design for a QIP device using a linear array of trapped electronic qubits on helium[9].

Surface-state electrons on liquid helium[10] are attracted by a weak positive image charge in the liquid and are held by a vertical pressing electric field $E_z$. This produces a vertical potential well with a series of excited states, similar to the Rydberg states in a hydrogen atom. Below 2 K, the electrons are in the quantum ground state, "floating" about 11 nm above the surface. Excited Rydberg states can be populated by microwave absorption at frequencies above 125 GHz[11]. An electronic qubit[9] would have the ground state as |0> and the first excited state as |1>. The electrons are well decoupled from the environment, interacting only weakly with thermal vibrations, or ripplons, on the atomically smooth superfluid helium surface and can have very high mobilities. Electron-electron Coulomb interactions are also important and a 2-dimensional (2D) electron Wigner crystal forms at low temperatures[10]. These unique properties make electrons on helium excellent candidates for qubits, with long decoherence times[9]. An essential requirement is to trap and control individual localised electrons, as reported here.

The new devices[12] are shown in Fig. 1. Free electrons are generated by thermionic emission from a pulsed filament. The electrons are stored on the helium surface in an electron reservoir, Fig. 1(a), which consists of a 10 μm wide helium microchannel[13], $d =$ 0.8 μm deep, above an electrode positively biased at $V_R$, which also extends into the circular helium pool whose surface forms an electron trap. This injector electrode enables electrons to be transferred between the reservoir and the electron trap, controlled by d.c. voltages on the SET, gate and reservoir electrodes. A single-electron transistor (SET) is placed beneath the helium surface in the electron trap. Periodic Coulomb blockade oscillations (CBO) are observed in the d.c. current through the SET, as shown in Fig. 2(a), as the gate voltage $V_g$ is swept. Each oscillation corresponds to an extra electronic charge $-e$ induced in the SET island, $Q_c = -C_{g1}V_g$, by capacitive coupling $C_{g1}$ from the gate electrode (note that a positive gate voltage induces negative charge in the SET). The CBO period $\Delta V_{g1} = 7.3$ mV, corresponds to $C_{g1} = e/\Delta V_{g1} = 21.92$ aF. For an uncharged helium pool the relative long-term charge stability of the SET is about $0.01e$. The linear variation of $Q_c$ with $V_g$ provides a reference for the extra induced charge $\Delta Q^* = Q^* - Q_c = Q^* + C_{g1}V_g$ from surface-state electrons on the helium. Each surface-state electron in the trap induces a *positive* fractional charge in the SET island

$$\frac{\Delta Q_1^*}{e} = \frac{c_1}{c_1 + c_2} > 0 \qquad [1]$$

where $c_1$ and $c_2$ are the capacitive electrostatic couplings from the free charge to the island and the rest of the world respectively. This changes the phase $\phi$ of the CBO by $\Delta\phi = 2\pi(\Delta Q_1^*/e)$ which can be measured, using the SET as a very sensitive electrometer.

The CBOs for a charged pool are shown in Fig. 2(a). As $V_g$ is swept, jumps in the CBO phase ϕ are observed, relative to the oscillations for an uncharged pool. These correspond to positive steps in $\Delta Q^*/e$ above the uncharged baseline, as electrons enter or leave the electron trap. In the example shown, some 5 electrons are counted leaving the trap. Free electrons are attracted into the electron trap by a positive gate potential, inducing a positive charge on the SET island and giving a *negative* differential capacitance. This picture is seen only after charging the helium surface. If the steps were produced by the movement of charge within the SET or on the dielectric surfaces, then they would be in the same direction as for capacitive coupling. Random phase shifts and two-level fluctuators[14] are sometimes observed from the movement of substrate charges, but with different characteristics from charging the electron trap. By sweeping the electrodes negative we can remove all the surface-state electrons and recover the uncharged results.

The steps in Fig. 2 reflect the discrete nature of electronic charge. The voltage, or Coulomb gap, required to increase the charge on a capacitor $C$ by one electron is $\Delta V = e/(C + V(dC/dV)) = e/C^*$, allowing for variations in $C$ with voltage $V$ (i.e. from changes in the electron trap size). If the gate electrode is coupled to the trapped electrons through a capacitance $C_{g2}$ then a gate voltage increment $\Delta V_{g2} = e/C^*_{g2}$ is required to attract each extra electron into the trap. The data in Fig. 2(a) gives $C^*_{g2} \approx 13$ aF. The size of the charge steps $\Delta Q^*_1/e$ is typically 0.092 $e$, depending on the electrode potentials. An electrostatic model has been developed for this electron trap that reproduces these values and demonstrates the effects of the gate and reservoir potentials. Fig. 2c shows model potential energy plots for emptying a 5-electron trap, controlled by the gate. A simple estimate of the number of electrons $N$ in a 2D trap is obtained from $N \approx CV(R)/e$, where $C = 8\varepsilon_0 R$ is the capacitance of an unscreened charged pool of radius $R$, and $V(R)$ is the potential in the pool, with $R$ and $V(R)$ measured from the potential minimum. A 10 mV trap, of radius 1 μm, will hold $N \approx 4$ electrons.

Fig. 3(a) shows charging the trap by sweeping the reservoir voltage $V_R$, starting from a large positive voltage (point A) where there is no potential barrier between the reservoir and the electron trap (see Fig.3(b)). The voltage of the surface-state electrons in the reservoir will be $V_{el} \approx V_R - ned/\varepsilon\varepsilon_0 = V_R - V_n$ where $n$ m$^{-2}$ is the number density of the electrons on the reservoir. As $V_R$ is reduced, a potential barrier forms. At point B the induced charge $Q^*$ on the SET island increases rapidly as electrons spill over from the reservoir to fill the trap. This continues until $V_{el} = 0$ (or $V_R = V_n$) when electrons will no longer be confined by the grounded guard electrode. When $V_R$ is now swept positive, electrons remain in the trap. At C the potential of the electrons in the reservoir falls below the barrier and the trapped electrons become isolated. As $V_R$ increases further the barrier decreases and electrons escape from the trap one at a time, giving a series of small steps in $\Delta Q^*/e \approx 0.1$.

At low temperatures, as in this experiment, Coulomb interactions between the electrons in the trap leads to localised electrons in specific structural arrangements, related to the 2D Wigner crystal or a Coulomb glass, depending on the strength of any disorder potential. In this region, the simple Coulomb gap capacitance model will be an approximation. This has been studied in numerical simulations and a wide range of phenomena have been predicted. Bedanov and Peeters[15] calculated the charging energies for equilibrium trapped electrons in a 2D parabolic potential well. It has been shown[16] that a metallic screening electrode close to a 2D electron sheet, as in our device, can lead to polaronic effects, in which the electronic spatial order is rearranged after the addition of extra electrons. One specific prediction was for simultaneous multiple electron addition, as seen in some semiconducting quantum dots[17]. The charging sequences in Fig.

4(a) show (i) double electron addition and (ii) hysteresis, consistent with these polaronic ideas. The distribution of the voltage charging increments (the addition spectrum) in the hysteretic regime is shown in Fig. 4b, in good agreement with the theoretical expression for a localised 2D Coulomb system with disorder[18]. The experimental disorder implied by this could arise from the electron injection dynamics (i.e. how the electrons re-arrange themselves in the trap) or from small variations in the trap potentials from substrate charges or surface-state electrons on the guard or reservoir.

Fig. 5(a) shows a single electron in a quantum well trap, as the first in a charging sequence. This enables a novel design for a linear chain quantum information processor[19], Fig. 5(b). An electron channel reservoir feeds electrons across an SET detector into a series of single-electron traps as qubits, Fig. 5(c), controlled by electrodes and microwave pulses. The crucial read-out stage, following a processing sequence, starts by ionising those electrons in the upper $|1>$ state. The remaining $|0>$ electrons are then conveyed along the linear trap array and detected with the SET in classical time, hence reading the output register.

To conclude, we have shown that we can trap electrons in a microfabricated potential well and detect and count them with a single-electron transistor (SET). The detection of individual localised electrons is of great interest in itself and is an essential precursor to the detection of the quantum state of individual electrons and their potential application as qubits.

We thank A.J.Dahm, M.I.Dykman, J.Goodkind, P.J.Meeson and J.Saunders for discussions and F.Greenough, A.K.Betts and others for technical support. The work was supported by the EPSRC, by the EU Human Potential Programme under contract HPRN-CT-2000-00157 Surface Electrons, and by Royal Holloway, University of London.

FIG. 1. The new microelectronic devices, fabricated on a Si/SiO2 substrate, by depositing Al, Au and Nb electrodes[12]. (a) A Nb guard electrode defines an electron reservoir and a helium pool (the electron trap), 0.8 μm deep, filled with superfluid helium. (b) Photograph of the 5 μm diameter electron trap, showing the electrometer (SET island) and the injector and gate electrodes. The SET is fabricated using shadow evaporation. Aluminium source (S) and drain (D) electrodes are connected to the SET island through Al/A$_2$O$_3$/Al tunnel barriers. It is operated below 0.5 K in the superconducting state at the Josephson-quasiparticle (JQP) peak in the I-V characteristic, with a source-drain voltage bias of 0.55 mV and a source-drain current $I_{SD} \approx 5$ nA. (c) Electrostatic model showing an electronic trap between a repulsive gate potential and an attractive electron reservoir potential.

FIG. 2. Counting individual electrons. (a) Coulomb blockade oscillations (CBO) in $I_{SD}$ for an uncharged (dashed line) and a charged (solid line) helium pool at 150 mK. The charge induced in the SET island, $Q^* = (\phi/2\pi)e$, where $\phi$ is the phase of the CBO, relative to a reference point. The arrows show phase and charge steps for the charged pool, giving an offset in the CBO compared to the uncharged state. (b) Charge steps $\Delta Q^*/e = (Q^* + C_{g1}V_g)/e$ from Fig. 2(a), as individual electrons enter the trap as $V_g$ is swept, with SET biased at $V_{SET} = 13.5$ mV. The dashed lines are spaced at intervals of $0.092e$. Inset: Model potential energy of a 5-electron trap for $V_g = +40, 0$ and $-40$ mV, showing inversion of the potential well for negative gate voltages. The model suggests that the effective contact potentials between the **different metals are small and that screening electrons may be trapped above the SET source and drain electrodes.**

**FIG. 3. Filling the electron trap. (a) Induced charge $\Delta Q^*/e$ while sweeping the reservoir voltage $V_R$ for charged and uncharged helium. Arrows show steps in induced charge as the trap empties. (b) Model potential energy of the electron trap for $V_R$ = 10, 120 and 240 mV, showing the reduction in the barrier height as $V_R$ increases.**

**FIG. 4. Charging energies. (a) Double electron steps and hysteresis between up (□) and down (o) gate voltage sweeps. The dashed lines are spaced at intervals of 0.092$e$. (b) Probability of a charge step per mV versus the voltage increment $\Delta V_g$ between the charge steps in Fig. 4(a). The mean voltage increment $\Delta V_{av}$ = 11.8 mV. The solid line shows the theoretical probability distribution $F(x) = 4(x-x_0)\exp[-2(x-x_0)^2]$ for a disordered system[18], for $x = \Delta V_g /V_{av} > 0.37 = x_0$.**

**FIG. 5. Trapping single electrons. (a) The trapping of a single localised electron (named Eddy) is shown, as the gate voltage is swept with $V_{SET}$ = 42 mV. The solid line shows the fitted theoretical voltage steps for the discrete charging of a parabolic potential well[15]. (b) Schematic cross-sections of a linear quantum information processor in a helium microchannel showing the electron reservoir, electronic qubits, control electrodes and SET detector read-out. The qubits are held in periodic potential wells. (c) Potential energy profile.**

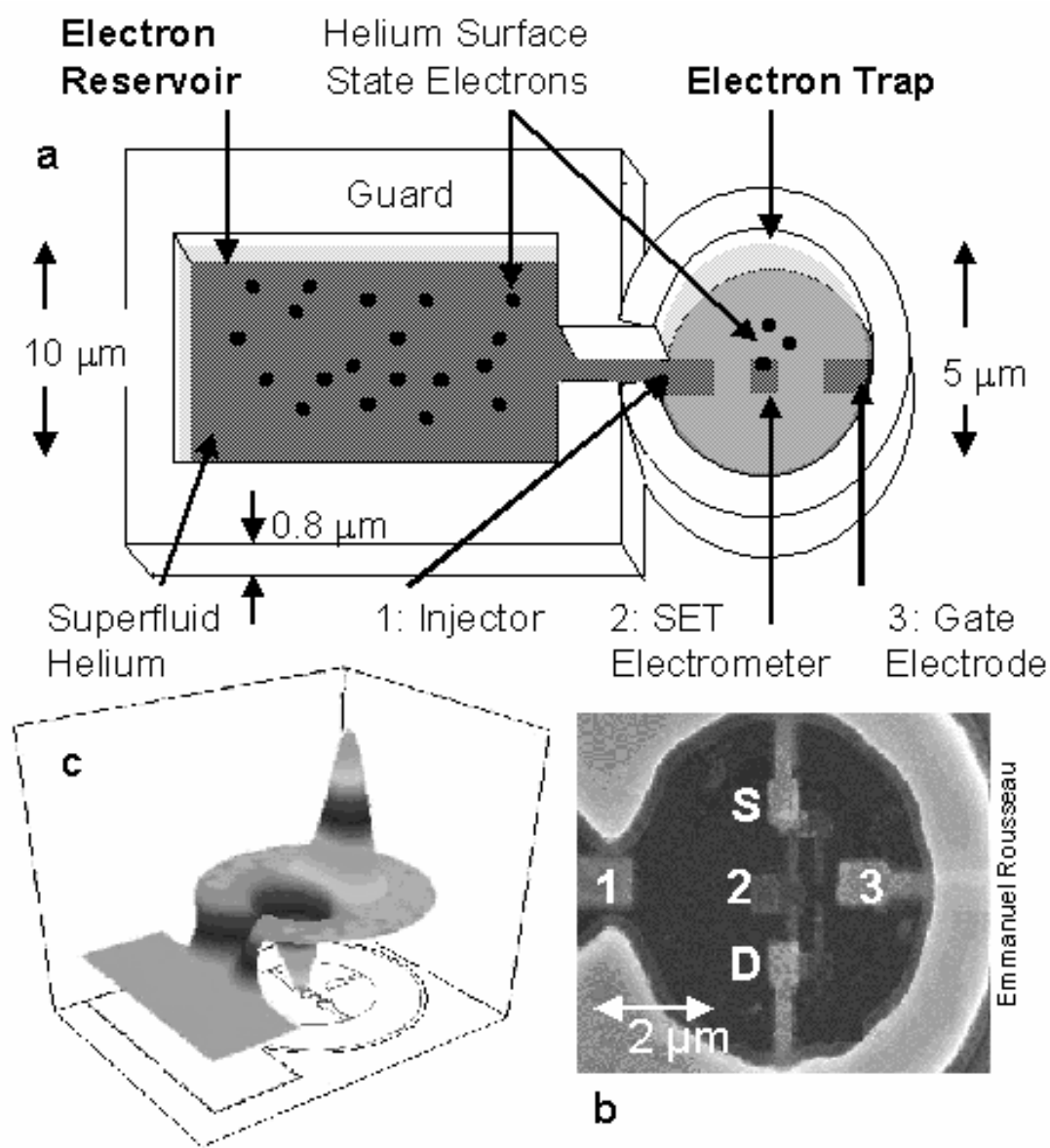

Fig.1

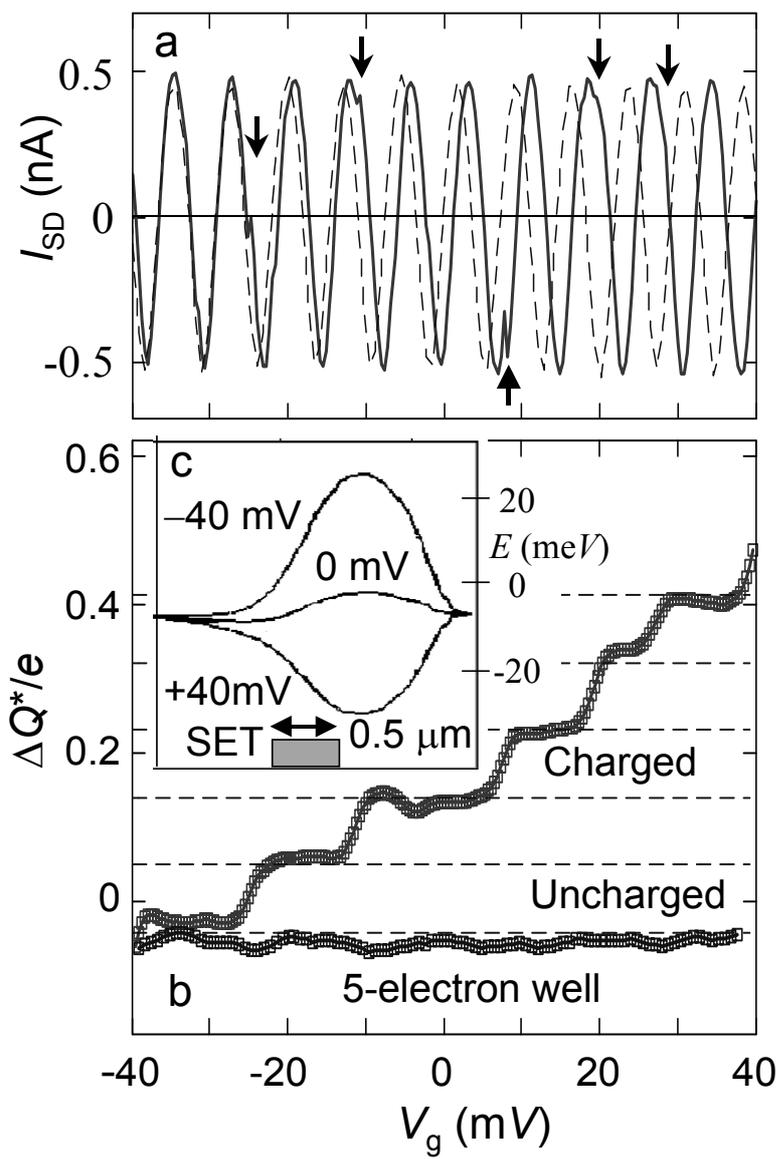

Fig.2

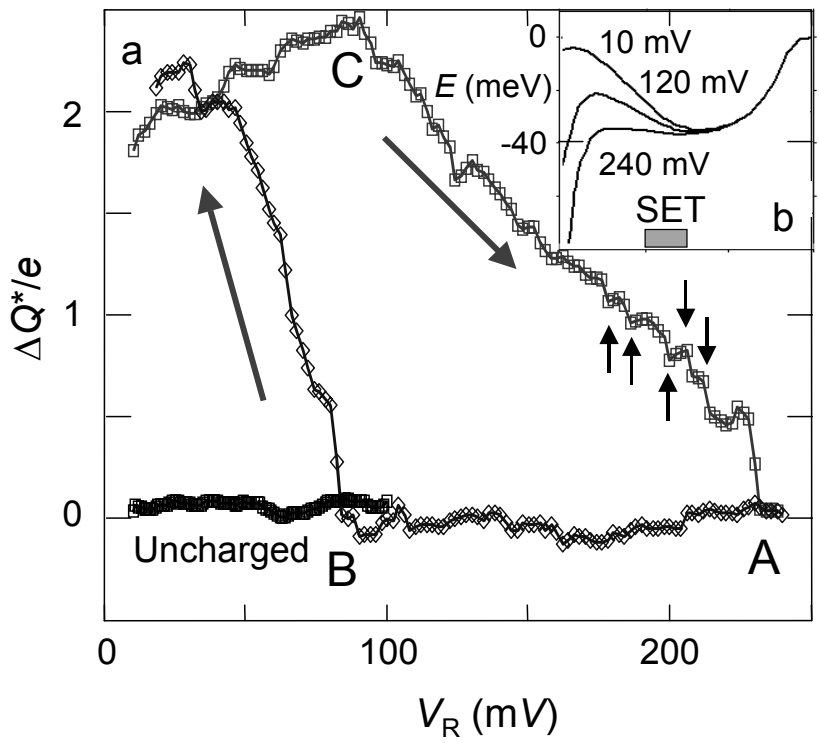

Fig.3

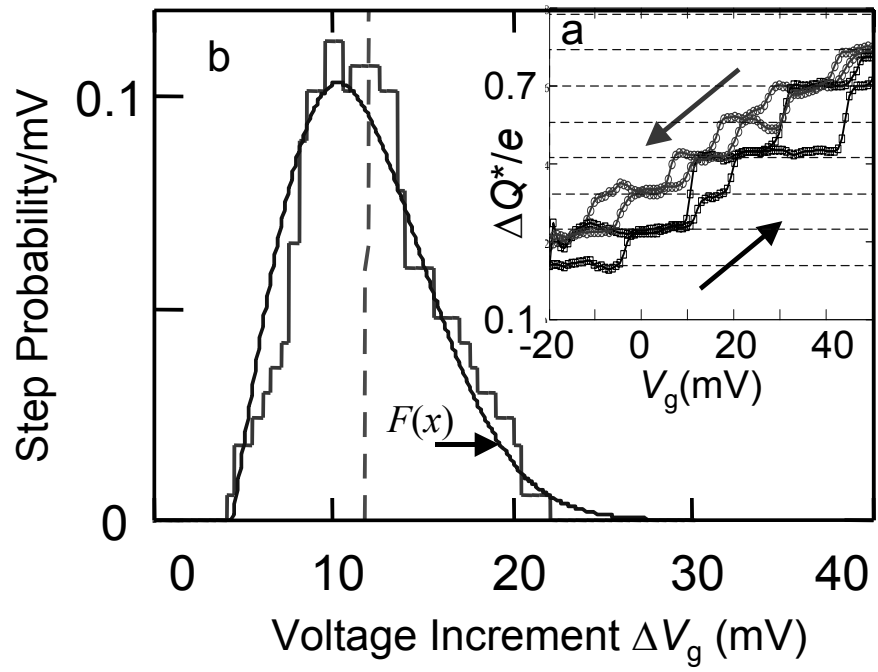

Fig.4

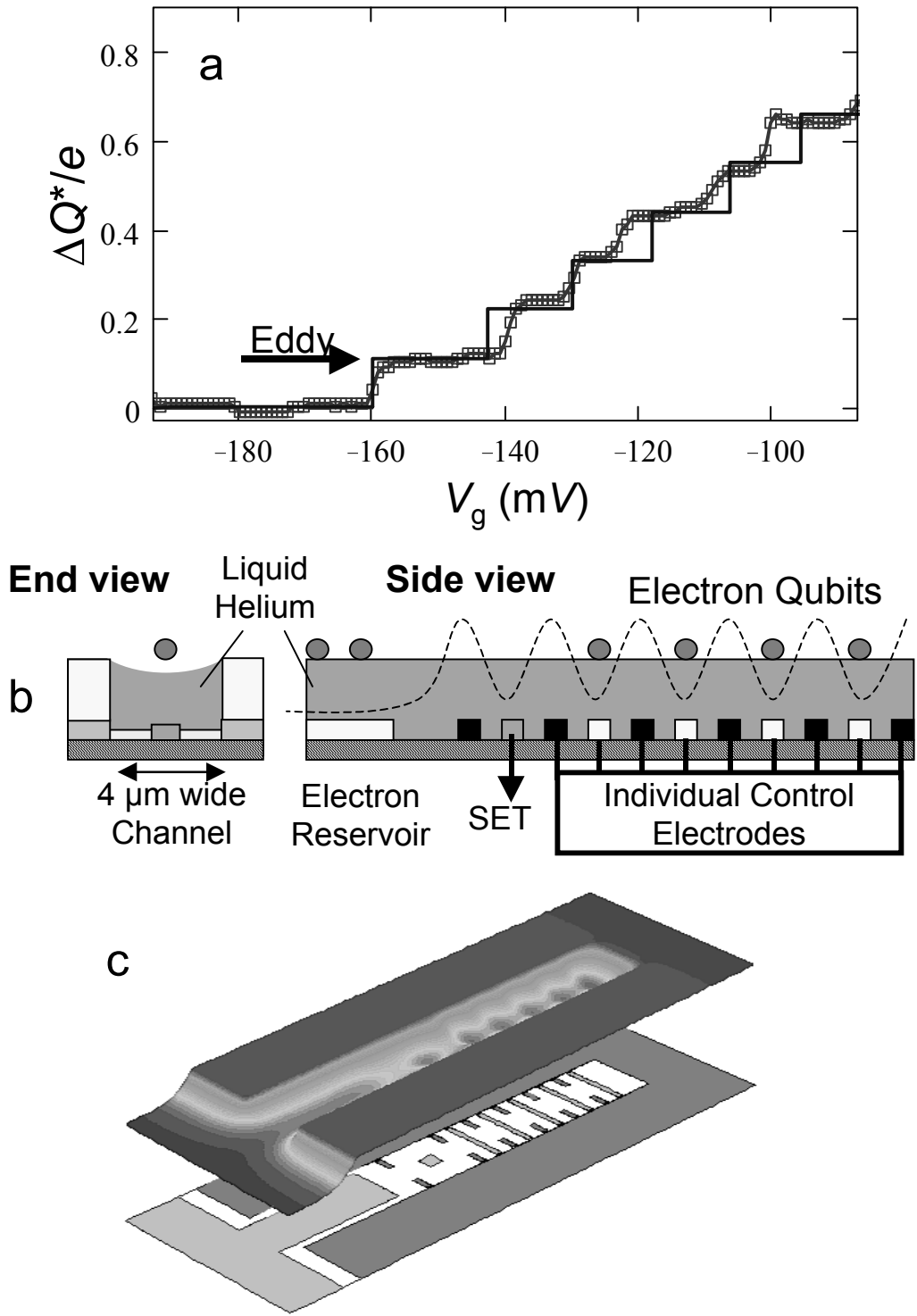

Fig.5